\documentclass[10pt]{article}

\usepackage{amsmath}
\usepackage{amssymb}

\usepackage{graphicx}

\usepackage{cite}

\usepackage{color} 

\usepackage[labelfont=bf,labelsep=period,justification=raggedright]{caption}

\makeatletter
\renewcommand{\@biblabel}[1]{\quad#1.}
\makeatother

\date{}

\begin{document}

\begin{flushleft}
{\Large
\textbf{Learning of Precise Spike Times with Homeostatic Membrane Potential Dependent Synaptic Plasticity}
}
\\
Christian Albers$^{1,\ast}$, 
Maren Westkott$^{1}$, 
Klaus Pawelzik$^{1}$
\\
\bf{1} Institute for Theoretical Physics, University of Bremen, Germany
\\
$\ast$ E-mail: calbers@neuro.uni-bremen.de
\end{flushleft}

\section*{Abstract}

Precise spatio-temporal patterns of neuronal action potentials underly e.g.
sensory representations and control of muscle activities. However, it is not
known how the synaptic efficacies in the neuronal networks of the brain adapt
such that they can reliably generate spikes at specific points in time.
Existing activity-dependent plasticity rules like Spike-Timing-Dependent
Plasticity are agnostic to the goal of learning spike times. On the other hand,
the existing formal and supervised learning algorithms perform a temporally
precise comparison of projected activity with the target, but there is no known
biologically plausible implementation of this comparison. Here, we propose a
simple and local unsupervised synaptic plasticity mechanism that is derived
from the requirement of a balanced membrane potential. Since the relevant
signal for synaptic change is the postsynaptic voltage rather than spike times,
we call the plasticity rule Membrane Potential Dependent Plasticity (MPDP).
Combining our plasticity mechanism with spike after-hyperpolarization causes a
sensitivity of synaptic change to pre- and postsynaptic spike times which can
reproduce Hebbian spike timing dependent plasticity for inhibitory synapses as
was found in experiments. In addition, the sensitivity of MPDP to the time course
of the voltage when generating a spike allows MPDP to distinguish between weak (spurious) and
strong (teacher) spikes, which therefore provides a neuronal basis for the
comparison of actual and target activity. For spatio-temporal input spike
patterns our conceptually simple plasticity rule achieves a surprisingly high
storage capacity for spike associations. The sensitivity of the MPDP to the
subthreshold membrane potential during training allows robust memory retrieval
after learning even in the presence of activity corrupted by noise. We propose that
MPDP represents a biologically realistic mechanism to learn temporal target
activity patterns.

\section*{Introduction}

Precise and recurring spatio-temporal patterns of action potentials are
observed in various biological neuronal networks. In zebra finches, 
precise sequences of activations in region HVC are found during singing and
listening to the own song~\cite{Hahnloser2002}. Also, when spike times of
sensory neurons are measured, the variability of latencies relative to the
onset of a externally induced stimulus is often higher than if the latencies are
measured relative to other sensory neurons~\cite{Gollisch2008, Masquelier2013};
spike times covary. Therefore, information about the stimulus is coded
in spatio-temporal spike patterns. Theoretical considerations show that in
some situations spike-time coding is superior to rate coding
~\cite{vanRullen2001}. Xu and colleagues demonstrated that through associative
training it is possible to imprint new sequences of activations in visual
cortex~\cite{Xu2012}, which shows that there are plasticity mechanisms which
are used to learn precise sequences.

These observations suggest that spatio-temporal patterns of spike activities
underlie coding and processing of information in many networks of the
brain. However, it is not known which synaptic plasticity mechanisms enable
neuronal networks to learn, generate, and read out precise action potential
patterns. A theoretical framework to investigate this question is the
chronotron, where the postsynaptic neuron is trained to fire a spike at
predefined times relative to the onset of a fixed input pattern
~\cite{Florian2012}. A natural candidate plasticity rule for chronotron training
is Spike-Timing Dependent Plasticity (STDP)~\cite{Caporale2008} in combination
with a supervisor who enforces spikes at the desired times. Legenstein and
colleagues~\cite{Legenstein2005} investigated the capabilities of supervised
STDP in the chronotron task and identified a key problem: STDP has no means to
distinguish between desired spikes caused by the supervisor and spurious spikes
resulting from the neuronal dynamics. As a result every spike gets reinforced,
and plasticity does not terminate when the correct output is achieved, which
eventually unlearns the desired synaptic state. The failings of STDP hint at the
requirements of a working learning algorithm. Information about the type of a
spike (desired or spurious) has to be available to each synapse, where it modulates
spike time based synaptic plasticity. Synapses evoking undesired spikes should
be weakened, synapses that contribute to desired spikes should be
strengthened, but only until the self-generated output activity matches the desired
one. Plasticity should cease if the output neurons generate the desired spikes
without supervisor intervention. In other words, at the core of a learning
algorithm has to be a comparison of actual and target activity, and synaptic
changes have to be computed based on the difference between the two.

In recent years, a number of supervised learning rules have been proposed to
train to fire temporally precise output spikes in response to recurring
spatio-temporal input patterns~\cite{Bohte2002, Florian2012, Memmesheimer2014}.
They compare the target spike train to the self-generated (actual) output and devise synaptic
changes to transform the latter into the former. However, because spikes are
discrete events in time that influence the future dynamics of the neuron, the comparison is necessarily non-local in time, which
might be difficult to implement for a biological neuron and synapse. Another
group of algorithms performs a comparison of actual and target firing rate
instead of spike times~\cite{Ponulak2010, Xie2004, Brea2013, Urbanczik2014}. Because they
work with the instaneous firing rate, they do not rely on sampling of discrete
spikes and therefore the comparison is local in time. It is interesting to note 
that these learning algorithms are implicitely sensitive to the current
membrane potential, of which the firing rate is a monotonous function. However,
two important questions remain unanswered: How is the desired activity communicated to a
biological neuron and how does the synapse compute the difference?

In this study, we investigate the learning capabilities of a plasticity rule
which relies only on postsynaptic membrane potential and presynaptic spikes as
signals. To distinguish it from spike times based rules, we call it Membrane
Potential Dependent Plasticity (MPDP). We derive MPDP from a homeostatic
requirement on the voltage and show that in combination with spike
after-hyperpolarisation (SAHP) it is compatible with experimentally observed
STDP of inhibitory synapses~\cite{Haas2006}. Despite its Anti-Hebbian nature,
MPDP combined with SAHP can be used to train a neuron to generate desired
temporally structured spiking output in an associative manner. During learning, the supervisor 
or teacher induces spikes at the desired times by a strong input. Because of
the differences in the time course of the voltage, a synapse can sense the
difference between spurious spikes caused by weak inputs and teacher spikes caused by strong inputs.
As a consequence, weight changes are matched to the respective spike type. Therefore,
our learning algorithm provides a biologically plausible answer for the open question
presented above. Additionally, the sensitivity of MPDP to subthreshold voltage leads to a
noise-tolerant network after training with noise free examples. For a
quantitative analysis, we simplify the neuron model and apply our learning
mechanism to train a Chronotron~\cite{Florian2012}. We find that the attainable
memory capacity is comparable to that of a range of existing learning rules
~\cite{Ponulak2010, Florian2012, Memmesheimer2014}, however the noise tolerance
after training is superior in networks trained with MPDP in comparison to those
trained with the other learning algorithms. 

\section*{Materials and Methods}

In this section, we present the models used. We start with the simpler leaky
integrate-and-fire neuron model (LIF neuron) and use it to derive the MPDP
rule. We then show how MPDP can be applied to a more realistic conductance
based integrate-and-fire neuron. Next, we describe the Chronotron setup we use
for quantitatively assessing the memory capacity of MPDP. Last, we provide the
definitions of the learning algorithms we used to compare MPDP with.

\subsection*{The LIF neuron and derivation of MPDP}

We investigated plasticity processes in a simple single-layered feed-forward
network with $N$ (presynaptic) input neurons and one (postsynaptic) output
neuron (see Fig.~\ref{fig1} A). For the input population
we stochastically generate spatio-temporal spike patterns which are kept fixed
throughout training (frozen noise). We denote the time of the $k$-th spike of
presynaptic neuron with index $i$ as $t_i^k$.

The postsynaptic neuron is modelled as a LIF neuron. The evolution of the
voltage $V(t)$ over time is given by
\begin{equation}
 \tau_m \dot{V} = -V + I_{syn} + I_{ext} \ .
 \label{eq:LIF}
\end{equation}
$I_{syn}$ and $I_{ext}$ are synaptic and external currents, respectively, and
$\tau_m$ is the membrane time constant of the neuron. If
the voltage reaches the firing threshold $V_{thr}$ at time $t_{post}$, the
neuron generates a spike and undergoes immediate reset to the reset potential
$V_{reset} < 0$. In the absence of any input currents, the neuron relaxes to an
equilibrium potential of $V_{eq} = 0$. Synaptic currents are given by
\begin{equation}
 \tau_s \dot{I}_{syn} = -I_{syn} + \sum\limits_i w_i \sum\limits_k \delta \left(t - t_i^k \right) \ .
\end{equation}
$\tau_s$ is the decay time constant of synaptic currents and $w_i$ is the
synaptic weight of presynaptic neuron $i$. For ease of derivation of MPDP, we
reformulated the LIF model. Because of the
linearity of Eq.~\ref{eq:LIF}, we can write the voltage as the sum of
kernels for postsynaptic potentials (PSPs) $\varepsilon (s)$ and resets $R(s)$:
\begin{equation}
 V(t) = \sum\limits_i w_i \sum\limits_k \varepsilon (t - t_k^i) + \sum\limits_{t_{post}} R (t - t_{post}) + \int\limits_0^{\infty} \kappa (t - s)  I_{ext} (s) ds \ .
 \label{eqn:LIF_asSRM}
\end{equation}
$\kappa = \exp \left( - (t - s)/\tau_m \right)$ is the passive response kernel
by which external currents are filtered. The other kernels are
\begin{equation}
 \begin{split} \varepsilon(s) & = \Theta(s) \frac{1}{\tau_m - \tau_s} \left( \exp(-s/\tau_m) - \exp(-s/\tau_s) \right) \\
  R(s) & = \Theta (s) (V_{reset} - V_{thr}) \exp (-s/\tau_m) \ . \end{split} 
\end{equation}
$\Theta(s)$ is the Heaviside step function. This formulation is also known as
the simple Spike Response Model (SRM${}_0$,~\cite{Gerstner2002}).

We next derive the plasticity rule from the naive demand of a balanced membrane potential:
The neuron should not be hyperpolarized nor too strongly depolarized. This is a sensible
demand for the dynamics of a neuronal network, because it holds the neurons at sensitive working points and also keeps
metabolic costs down. For the formalization of the objective, we introduce an
error function which assigns a value to the current voltage:
\begin{equation}
 2 E(t) = \gamma \left( \left[ V(t) - \vartheta_D \right]_+ \right)^{2} + \left( \left[ \vartheta_P - V(t) \right]_+ \right)^{2} \ ,
 \label{eq:errorfunction}
\end{equation}
where $\vartheta_{D, P}$ are thresholds for plasticity, and $\gamma$ is a
factor that scales synaptic long-term depression (LTD) and long-term
potentiation (LTP) relative to each other. Whenever $V(t) > \vartheta_D$ or
$V(t) < \vartheta_P$, the error function is greater than zero. Therefore, to minimize the error,
the voltage must stay between both thresholds. In this study, we choose
$\vartheta_P = V_{eq}$.
$\vartheta_D$ is set between the firing threshold and $V_{eq}$. From the error
function, a weight change rule can be obtained by computing the partial
derivative of $E(t)$ with respect to weight $w_i$:
\begin{equation}
 \begin{split} \frac{\partial E(t)}{\partial w_i} & = 
  \gamma \left[ V(t) - \vartheta_D \right]_+ \frac{\partial V(t)}{\partial w_i} - \left[ \vartheta_P - V(t) \right]_+ \frac{\partial V(t)}{\partial w_i} \\
  & = \left( \gamma \left[ V(t) - \vartheta_D \right]_+  - \left[ \vartheta_P - V(t) \right]_+ \right) \sum\limits_k \varepsilon \left(t - t_i^k \right) \ . \end{split}
  \label{eq:dEdwi}
\end{equation}
The MPDP rule then reads
\begin{equation}
 \dot{w}_i =- \eta \frac{\partial E(t)}{\partial w_i} = 
 \eta \left(- \gamma \left[ V(t) - \vartheta_D \right]_+ + \left[ \vartheta_P - V(t) \right]_+ \right) \sum\limits_k \varepsilon \left(t - t_i^k \right) \ .
 \label{eq:MPDP}
\end{equation}
$\eta$ is the learning rate. The weights change along the gradient of the error
function, i.e. MPDP is a gradient descent rule that minimizes the error 
resulting from a given input pattern.

\subsection*{The conductance based LIF neuron}

The simple model above suffers from the fact MPDP is agnostic to the type of synapse.
In principle, MPDP can turn excitatory synapses into inhibitory ones by
changing the sign of any weight $w_i$. Since this is a violation of Dale's law,
we present a more biologically realistic scenario involving MPDP.
We split the presynaptic population into $N_e$ excitatory and $N_i$ inhibitory
neurons. The postsynaptic neuron is modelled as a conductance based LIF neuron
governed by
\begin{equation}
 C_m \frac{\text{d}V}{\text{d}t} = -g_L(V - V_L) - (g_{s} + g_{f})(V - V_h) - g_{ex}(V - V_{ex}) - g_{in}(V - V{in}) \ ,
 \label{eq:conductance_based_LIF}
\end{equation}
where $V$ denotes the membrane potential, $C_m$ the membrane capacitance, $V_L$
the resting potential, $g_L$ the leak conductance, $V_i$ and $V_{ex}$ the
reversal potential of inhibition and excitation, respectively and $g_{in}$ and
$g_{ex}$ their respective conductances. The spike after-hyperpolarisation is
modeled to be biphasic consisting of a fast and a slow part, described by
conductances $g_{f}$ and $g_{s}$ that keep the membrane potential close to
the hyperpolarisation potential $V_h$. When the membrane potential surpasses
the spiking threshhold $V_{thr}$ at time $t_{post}$, a spike is registered and
the membrane potential is reset to $V_{reset}=V_h$. All conductances are
modeled as step and decay functions. The reset conductances are given by
\begin{equation}
 \tau_{f, s} \dot{g}_{f, s} = - g_{f, s} + \Delta g_{f, s} \sum\limits_{t_{post}} \delta \left( t - t_{post} \right) \ ,
\end{equation}
where $\Delta g_{f, s}$ is the increase of the fast and slow conductance at
the time of each postsynaptic spike, respectively. They decay back with time
constants $\tau_{f} < \tau_{s}$. The input conductances
$g_{ex}$ and $g_{in}$ are step and decay functions as well, that are increased
by $w_i$ when presynaptic neuron $i$ spikes and decay with time constant
$\tau_s$. $w_i$ denotes the strength of synapse $i$.

In this model, we employ MPDP as defined by Eq.~\ref{eq:MPDP} with the
following restrictions:
\begin{itemize}
 \item Technically, there is no fixed PSP kernel for the conductance based model,
 since the amplitude of a single PSP depends on the current voltage. Still, we
 use the same rule by keeping track of ``virtual PSPs'' for each synapse that
 do not affect the neuronal dynamics.
 \item MPDP affects only inhibitory synapses. Excitatory ones are kept fixed.
 \item Because inhibitory synapses have negative impact on the neuron, we
 exchange LTP and LTD in the MPDP rule to account for that. Formally, we
 introduce thresholds $\vartheta_D^I$ and $\vartheta_P^I$.
\end{itemize}
$\vartheta_D^I$ lies below the equilibrium potential $V_{L}$, and an inhibitory
synapse depresses whenever it is active and $V(t) < \vartheta_D^I$. Similarly,
when $V(t) > \vartheta_P^I$, any active inhibitory synapse gets potentiated.
Note that the qualitative effect on the membrane potential remains unchanged to the
example in Fig.~\ref{fig1} B.

\subsection*{Evaluation of memory capacity}

The memory capacity of a typical neuronal network in a given task crucially
depends on the learning rules applied (for an example in spiking networks see
~\cite{Florian2012}). Recently, it was shown that the maximal number of spiking
input-output associations learnable by a postsynaptic neuron lies in the range
of 0.1 to 0.3 per presynaptic input neuron~\cite{Memmesheimer2014}. The exact
number mostly depends on the shape of the PSP (determined by $\tau_m$ and
$\tau_s$) and to a lesser extent on average pre- and postsynaptic firing rates.
Here, we evaluate the memory capacity attainable with MPDP and
compare it with ReSuMe~\cite{Ponulak2010}, E-Learning~\cite{Florian2012} and
FP-Learning
~\cite{Memmesheimer2014}, with the latter learning rule being optimal in terms
of memory capacity. For ease of comparison, we adapt the Chronotron setting
introduced by Florian~\cite{Florian2012}, use the simple neuron model
of the LIF neuron and let synapses change their sign. The definitions of
patterns, associations and memory capacity is similar to the ones used in
Tempotron and Perceptron training~\cite{Gutig2006, HKP}. We provide a
short description of ReSuMe, E-Learning and FP-Learning below.

\subsubsection*{Chronotron setting}

The goal of the Chronotron is to imprint input-output associations into the
weights. 
One input pattern consists of spatio-temporal spiking activity of the $N$ input
neurons with duration $T = 200 ms$. In each pattern, each input neuron spikes
exactly once, with spike times $t_i^\mu$ drawn i.i.d. from the interval
$[0, T]$. $\mu \in 1, \dots, P$ indexes the patterns. For each input pattern we
draw one desired output spike time $t_d^{\mu}$ i.i.d. from the
interval $[\Delta_{edge}, T - \Delta_{edge}]$, with $\Delta_{edge} = 20 ms$. We reduce the length of the output interval
to ensure that each output spike in principle can be generated
by the input. If the desired output spike time is too early there might be
no input spikes earlier than $t_d$, which makes it impossible for the
postsynaptic neuron to generate the desired output. After all $P$
patterns have been generated, we keep them fixed for the duration of network
training and recall testing. Training is organized in learning trials and
learning blocks. A learning trial in MPDP consists of the presentation of one
of the input patterns and concurrent induction of a teacher spike at time
$t_d^{\mu}$ by injection of a supratheshold delta-pulse current by the
supervisor. In all other learning rules, the supervisor passively observes the
output activity and changes weights afterwards based on the actual output. A
learning block consists of $P$ learning trials, with each of the different
input patterns presented exactly once in randomized order. After each learning
trial, synaptic weights are updated. After each learning block, we present the
input patterns again to test the recall quality. Supervisor intervention and
synaptic plasticity are switched off for recall trials.

\subsubsection*{Computing the capacity}

We test the capacity of each learning rule (MPDP, ReSuMe, E-Learning and
FP-Learning) by training networks of different sizes, $N \in \{200, 500, 1000,
2000\}$. Because we assume that the number of patterns or input-output
associations that can be learned scales with $N$~\cite{Florian2012,
Memmesheimer2014}, we introduce the
load parameter $\alpha$ with $P = \alpha N$. We pick discrete $\alpha \in
[0.01, 0.3]$. For each combination of $\alpha$ and $N$, we generate 50
different realizations of $P$ patterns and $N$ initial weights, which are
drawn from a gaussian distribution with mean and standard deviation $T \cdot
30 mV / N$. For a non-spiking neuron (i.e. Eq.~\ref{eq:LIF} with $V_{thr}
\gg 1$) this would result in an average membrane potential of $30 mV$ before
learning. As a result initially the postsynaptic neuron fires several
spurious spikes. This way we test the ability of a learning rule to
extinguish them.

After each learning block, the recall is tested. Recall is counted as a success
if the postsynaptic neuron fires exactly one output spike in a window of length
$4 ms$ centered around $t_d^{\mu}$, and no additional spike at any
time. We define success loosely, because MPDP and FP-Learning do not
converge onto generating the output spike exactly at $t_d^{\mu}$.

We train each network for a fixed number of learning blocks (10000 in the case
of MPDP, 20000 for the others). Because we evaluate recall after each learning
block, we can check whether the system has converged. We define capacity as the
``critical load'' $\alpha_{90}$, where on average 90 \% of the spikes are
recalled after training. To approximate it, we plot the fraction of patterns
correctly learned as a function of the load $\alpha$. The critical load is defined as
the point where a horizontal line at 90 \% correct recall meets the graph.

\subsubsection*{Testing noise tolerance}

The threshold for LTD, $\vartheta_D$, is not only a way to impose homeostasis on
the synaptic weights. It is also a safeguard against spurious spikes that could
be caused by fluctuations in the input or membrane potential. The reason is
that after convergence of weight changes for known input patterns the voltage
mostly stays below $\vartheta_D$ for all non spike times due to the repulsion
of the membrane potential away from threshold. This leaves room for the voltage
to fluctuate without causing spurious spikes.

We apply noise in two conditions. First we want to know if a trained network is
able to recall the learned input-output associations in the presence of noise,
i.e. we train the network first and apply noise only during the recall trials.
Second we test if a learning rule can be used to train the network in the
presence of noise. In this condition, we test recall noise free.

We induce noise in two different ways. One way is to add a stochastic external
current
\begin{equation}
 I_{ext} (t) = \frac{\sigma_{input}}{\sqrt{\tau_m}} \eta (t) \ .
\end{equation}
$\eta (t)$ is a gaussian white noise process with zero mean and unit variance.
The factor makes sure that the actual noise on the membrane potential has
standard deviation $\sigma_{input}$.

The other way is to jitter the input spike times. Instead of using presynaptic
spike times $t_i^{\mu}$, we let the neurons spike at times
\begin{equation}
 \hat{t}_i^{\mu} = t_i^{\mu} + \mathcal{N} (0, \sigma_{jitter}) \ ,
\end{equation}
where $\mathcal{N} (0, \sigma)$ is a random number drawn from a gaussian
distribution with standard deviation $\sigma$.

If we apply noise only during recall, we use the weights after the final
learning block and for each noise level $\sigma_{input, jitter}$ we average the
recall over 50 seperate noise realizations and all training realizations.

Although both procedures lead to random fluctuations of the membrane potential
in each pattern presentation, they lead to different results. The reason is
that by using jitter on the input spike times, the statistics of the weights
impact on the actual amount of fluctuations of the voltage. This has noticable
consequences for the different learning rules. 

\subsection*{Learning algorithms used for quantitative comparison}

Our goal is a quantitative analysis of the memory capacity of MPDP in the
Chronotron task. We feel this necessitates a comparison to other learning
rules. We chose ReSuMe~\cite{Ponulak2010}, which is a prototypical
learning rule for spiking output, E-Learning~\cite{Florian2012} as a powerful
extension, and FP-Learning~\cite{Memmesheimer2014}, which was shown to achieve
optimal memory capacity in the task. Here, we provide a short description of all
three rules.

\subsubsection*{The $\delta$-rule and ReSuMe}

The $\delta$-rule, also called the Widrow-Hoff-rule~\cite{HKP}, lies at the core of a
whole class of learning rules used to teach a neuronal network some target
activity pattern. Synaptic changes are driven by the difference of desired and
actual output, weighted by the presynaptic activity:
\begin{equation}
 \Delta w (t) \propto f_{pre} (t) \left( f_{post}^{target} (t) - f_{post}^{actual} (t) \right) \ .
\label{eqn:WidrowHoff}
\end{equation}
We denote pre- and postsynaptic firing rate with $f_{pre, post}$. The
target activity $f_{post}^{target} (t)$ is some arbitrary time dependent firing
rate. The actual self-generated activity $f_{post}^{actual} (t)$ is given by
the current input or voltage of the postsynaptic neuron (depending on the
formulation), transformed by the input-output function $g(h)$ of the neuron.

ReSuMe (short for Remote Supervised Method) is a supervised spike-based
learning rule first proposed in 2005~\cite{Ponulak2010}. It is derived from the
Widrow-Hoff rule for rate-based neurons, applied to deterministic spiking
neurons. Therefore, continuous time dependent firing rates get replaced with
discrete spiking events in time, expressed as sums of delta-functions. Because
these functions have zero width in time, it is necessary to temporally spread
out presynaptic spikes by convolving the presynaptic spike train with a
temporal kernel. Although the choice of the kernel is free, usually a causal
exponential kernel works best. We also used ReSuMe with a PSP kernel to train
Chronotron, but the results were worse than with the exponential kernel. The
weight change is given by
\begin{equation}
 \dot{w} (t) \propto \left[ S_d (t) - S_o (t) \right] \left[ a_d + \int\limits_0^{\infty} \exp (-s/\tau_{plas} ) S_i(t - s) ds \right] \ ,
\end{equation}
where $S_d (t)$ is the desired, $S_o (t)$ is the self-generated and $S_i (t)$
the input spike train at synapse $i$. $\tau_{plas}$ is the decay time constant
of the exponential kernel. $a_d$ is a constant which makes sure
that the actual and target firing rates match; learning also works without,
therefore we choose $a_d = 0$ in our study. ReSuMe converges when both actual
and desired spike lie at the same time, because in this case the weight changes
cancel exactly.

In recent years, several rules for spiking neurons have been devised which are
similar to the $\delta$-rule~\cite{Xie2004, Brea2013, Urbanczik2014}. With the
``PSP sum''
\begin{equation}
 \lambda_i = \sum\limits_k \varepsilon (t - t_i^k) \ ,
\end{equation}
the weight change takes the form
\begin{equation}
 \dot{w}_i \propto \left[ S_{teacher} (t) - \rho (V(t)) \right] f(\rho(V(t))) \lambda_i (t) \ .
\end{equation}
$S_{teacher}$ is a stochastic realization of a given desired time dependent target
firing rate, $\rho (V(t))$ is the instantaneous firing rate, which depends on
the current membrane potential, and $f(\rho)$ is a function which
additionally scales the weight changes depending on the current firing rate.
Although the rule of Xie and Seung~\cite{Xie2004} was defined in a reward learning framework,
it is equivalent to the formulation above if the output neuron is forced to fire a
teacher spike train and reward is kept constant.

\subsubsection*{E-Learning}

E-Learning was conceived as an improved learning algorithm for spike time
learning~\cite{Florian2012}. It is derived from the Victor-Pupura distance
(VP distance) between spike trains~\cite{Victor1996}. The VP-distance is used
to compare the similarity between two different spike trains. Basically, spikes
can be shifted, deleted or inserted in order to transform one spike train
into the other. Each action is assigned a cost, and the VP distance is the
minimum transformation cost. The cost of shifting a spike is proportional to
the distance it is shifted and weighted with a parameter $\tau_q$. If the shift
is too far, it gets cheaper to delete and re-insert that spike.

E-Learning is a gradient descent on the VP distance and has smoother
convergence than ReSuMe. In this rule, first the actual output spike train is compared to the
desired spike train. With the VP algorithm it is determined if output spikes
must be shifted or erased or if some desired output spike has no close actual
spike so a new spike has to be inserted. Based on this evaluation, actual and
desired spikes are put in three categories:
\begin{itemize}
 \item{Actual output spikes are ``paired'' if they have a pendant, i.e. a
 desired spike close in time and no other actual output spike closer (and vice
 versa). These spikes are put into a set $S$.}
 \item{Unpaired actual output spike that need to be deleted are put into the
 set $D$.}
 \item{Unpaired desired output spike times are put into the set $J$, i.e. the
set of spikes that have to be inserted.}
\end{itemize}
To clarify, $S$ contains pairs of ``paired'' actual and desired spike times,
$D$ contains the times of all unpaired actual spikes, and $J$ the times of
unpaired desired spike times. With the PSP sum as above, the E-Learning rule is
then
\begin{equation}
 \Delta w_i = \gamma \left[ \sum\limits_{t_{ins} \in J} \lambda_i (t^{ins}) - \sum\limits_{t^{del} \in D} \lambda_i (t^{del}) + 
 \frac{\gamma_r}{\tau^2_q} \sum\limits_{(t^{act}, t^{des}) \in S} (t^{act} - t^{des}) \lambda_i (t^{act}) \right] \ .
\end{equation}
$\gamma$ is the learning rate, and $\gamma_r$ is a factor to scale spike
shifting relative to deletion and insertion.

The former two terms of the rule correspond to ReSuMe, except the kernel is not
a simple exponential decay. The advantage of E-Learning is that the weight
changes for spikes close to their desired location are scaled with the
distance, which improves convergence and consequentially memory capacity.

\subsubsection*{FP-Learning}

FP-Learning~\cite{Memmesheimer2014} was devised to remedy a central problem in
learning rules like ReSuMe and others. Any erroneous or missing spike
``distorts'' the time course of the membrane potential behind it compared to
the desired final state. This creates a wrong environment for the learning
rule, and weight changes can potentially be wrong. Therefore,
the FP-Learning algorithm stops the learning trial as soon as it encounters any
spike output error. Additionally, FP-Learning introduces a margin of tolerable
error for the desired output spikes. An actual output spike should be generated
in the window of tolerance $[t_d - \epsilon, t_d + \epsilon]$ with the
adjustable margin $\epsilon$. Weights are changed on two occasions:
\begin{enumerate}
 \item If a spike occurs outside the window of tolerance for any $t_d$ at time
 $t_{err}$, then weights are depressed by $\Delta w_i \propto - \lambda_i
 (t_{err})$. This also applies if the spike in question is the second one within
 a given tolerance  window.
 \item If $t = t_d + \varepsilon$ and no spike has occured in the window of
 tolerance, then $t_{err} = t_d + \varepsilon$ and $\Delta w_i \propto
 \lambda_i (t_{err})$.
\end{enumerate}
In both cases, the learning trial immediately ends, to prevent that the
``distorted'' membrane potential leads to spurious weight changes. Because of
this property, this rule is also referred to as ``First Error Learning''.

\subsection*{Parameters of the simulations}

\subsubsection*{Conductance based neuron}

The parameters used are as follows: $C_m = 0.25 nF$, $g_L = 20 nS$, $V_L = -55
mV$, $V_{thr}= - 50$ $V_{ex} = -40 mV$, $V_h = V_{reset} = V_{in}=-75mV$,
$\Delta g_{s} = 0.001$, $\Delta g_{f} = 0.04$, $\tau_{f} = 3 ms$, $\tau_{s}
= 12.5 ms$, and $\tau_s=3 ms$. For the MPDP rule, the parameters are:
$\vartheta_D^I = -58 mV$, $\vartheta_P^I = -53 mV$, $\gamma = 100$, $\eta =
5\cdot 10^{-8}$ and $\tau_m = C_m/g_L = 12.5 ms$.

\subsubsection*{Simple LIF neuron}

The neurons' parameters are $\tau_s = 3 ms$, $\tau_m = 10 ms$ and $V_{thr} = 20
mV$. The reset potential is $V_{reset} = -5 mV$ with MPDP and $V_{reset} = 0
mV$ for the other learning rules. For MPDP we use $\vartheta_D = 18 mV$,
$\vartheta_P = 0 mV$, $\gamma = 14$, and $\eta = 5*10^{-4}$. With ReSuMe, we
find $\tau_{plas} = 10 ms$, and $\eta = \{10, 4, 2, 1\} \cdot 10^{-10}$ for
200, 500, 1000 and 2000 neurons as good parameters. FP-Learning has only a
single free parameter, the learning rate $\eta = 10^{-9}$.

\subsubsection*{Numerical procedures}

All networks with MPDP were numerically integrated using a simple Euler integration
scheme. The simulations for the conductance based LIF neuron were written in
Python and used a step size of 0.01 ms. The neurons parameters are set to
values that are both biologically realistic and similar to those of the
quantitative analysis. For reference, we put them into the Supporting Informations.

The simulations of the simple neuron and scripts for analysis were written in
Matlab (Mathworks, Natick, MA). Here, we used a step size of 0.1 ms. 

The networks that were trained with ReSuMe, E-Learning and FP-Learning were
simulated using an event-based scheme~\cite{Dhaene2009}, since in these rules the subthreshold
voltage is not important.

The parameters like learning rates and thresholds we use are set by hand for
all plasticity rules. Before doing the final simulations, we did a search in
parameter space by hand to find combinations which yield high performance in
the Chronotron task.

The error we report in Fig.~\ref{fig4} C and D, Fig.~\ref{fig5} A to D and
Fig.~\ref{fig6} A and B is the standard error of the mean (SEM) over all 50
realizations.

\section*{Results}

In the following, we start with presenting our Membrane Potential Dependent
Plasticity rule (MPDP). We constructed a simple yet biologically plausible
feed-forward network and show that MPDP, when tested with spike pairs, is 
equivalent to inhibitory Hebbian STDP as
reported by Haas and colleagues~\cite{Haas2006}. We then show that with MPDP
the output neuron of this example can be trained to generate spikes at specific
times. Lastly, we turn to a simplified model to evaluate and compare with other
rules the attainable memory capacity with MPDP, as well as its noise tolerance.

\subsection*{Membrane Potential Dependent Plasticity}

\begin{figure}[!ht]
 \begin{center}
  \includegraphics[width=0.8\textwidth]{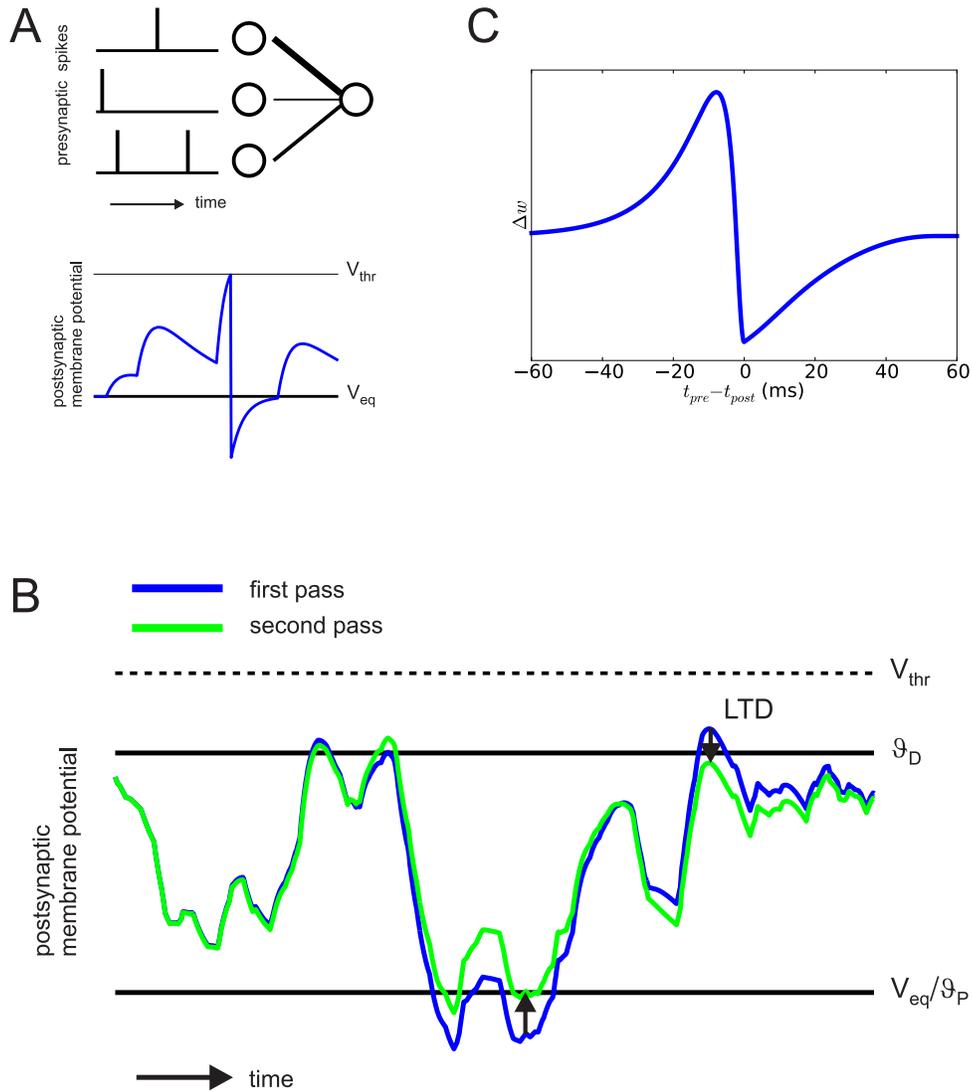}
 \end{center}
 \caption{{\bf A:} The model network has a simple feed-forward structure. The
 top picture shows three pre- and one postsynaptic neurons, connected by
 synapses. Line Width in this example corresponds to synaptic strength. Bottom
 picture shows the postsynaptic membrane potential in response to the input.
 {\bf B:} Illustration of Anti-Hebbian Membrane Potential Dependent
 Plasticity (MPDP). A LIF neuron is presented twice with the same presynaptic
 input pattern. Excitation never exceeds $V_{thr}$. MPDP changes synapses to
 counteract hyperpolarization and depolarization occuring in the first
 presentation (blue trace), reducing (arrows) them on the second presentation
 (green trace). {\bf C:} Homeostatic MPDP on inhibitory synapses is compatible
 with STDP as found in experiments. Plasticity is tested for different temporal
 distances between  pre- and postsynaptic spiking; the resulting spike timing
 characteristic is in  agreement with experimental data on STDP of inhibitory
 synapses~\cite{Haas2006}.}
 \label{fig1}
\end{figure}

We formulated a basic homeostatic requirement on the membrane
potential of a neuron. The neuron should stay in a sensible working regime; in
other words, its voltage should be confined to moderate values. We formalized
this by introducing two thresholds on the voltage. In this study, $\vartheta_D$ lies between the firing threshold and
resting potential and $\vartheta_P$ is equal to the resting
potential. With these thresholds, we formulated an error function (see Eq.~\ref{eq:errorfunction} in Methods). Using it and a simple LIF neuron model with
linear dynamics below the firing threshold, we computed an update rule for the
weights, Eq.~\ref{eq:MPDP}. Weight changes with this rule ``bend'' the
voltage at the times of non-zero error towards the region between the two
thresholds. Fig.~\ref{fig1} B shows how MPDP
effects the voltage for recurring input activity.

\subsection*{Homeostatic MPDP on inhibitory synapses is compatible with STDP}

We first investigated the biological plausibility of a network with MPDP.
Experimental studies on plasticity of cortical excitatory
neurons often find Hebbian plasticity rules like Hebbian
Spike Timing Dependent Plasticity (STDP; see~\cite{Markram1997, Feldman2000,
Sjostrom2001, Froemke2002, Wang2005} for examples). Reports on Anti-Hebbian plasticity or
sensitivity to subthreshold voltage in excitatory cortical neurons are scarce
~\cite{Sjostrom2004, Lamsa2007, Fino2009, Verhoog2013}. However, it has been
reported that plasticity in (certain) inhibitory synapses onto excitatory cells
has a Hebbian characteristic~\cite{Haas2006}, i.e. synapses active before a
postsynaptic spike become stronger, those active after the spike become weaker.
The net effect of this rule on the postsynaptic neuron is
\emph{Anti-Hebbian}, because weight increases tend to suppress output spikes.

In experimental investigations of STDP, neurons are tested with pairs of pre-
and postsynaptic spikes. We mimicked this procedure in a simple network
consisting of one pre- and one postsynaptic neuron, and one ``experimentator
neuron'' . The postsynaptic neuron was modelled as
a conductance based LIF neuron. The experimentator neuron has a fixed strong
excitatory synaptic weight onto the postsynaptic neuron, so that a spike of the experimentator neuron
causes a postsynaptic spike. We used it to control the postsynaptic spike
times. The presynaptic neuron is inhibitory and its weight is small compared to
the experimentator, so that it has negligible influence on the postsynaptic
spike time. We probed synaptic plasticity by inducing a pair of a pre- and a 
postsynaptic spike at times $t_{pre}$ and $t_{post}$, and vary $t_{pre}$ while
keeping $t_{post}$ fixed. The resulting weight change of the inhibitory neuron
as a function of timing difference is shown in Fig.~\ref{fig1} C. The shape
of the function is in qualitative agreement with experimental results~\cite{Haas2006}.

It is necessary to assume the presence of an ``experimentator
neuron''. The reason is that the shape of the STDP curve explicitely depends
on the specifics of spike induction since the MPDP rule is sensitive only to
subthreshold voltage. For example, using a delta-shaped input current would
lead to a LTD-only STDP curve, since the time the voltage needs to cross the
firing threshold starting from equilibrium is infinitely short.

\subsection*{Homeostatic MPDP allows associative learning}

At first glance, it might seem unlikely that a homeostatic plasticity mechanism
can implement associative learning. It is Anti-Hebbian in nature, because if
the membrane potential is close to firing threshold it gets suppressed, and
if is below the resting potential it gets lifted up. However, the neuronal dynamics shows
somewhat stereotypic behavior before, during and after each spike. To induce a spike,
the neuron needs to be depolarized up to $V_{thr}$, where active feed-back processes kick in.
These processes cause a very short and strong depolarization and a subsequent
undershoot of the membrane potential (hyperpolarization), from where it relaxes
back to equilibrium.

\begin{figure}[!ht]
 \begin{center}
  \includegraphics[width=0.95\textwidth]{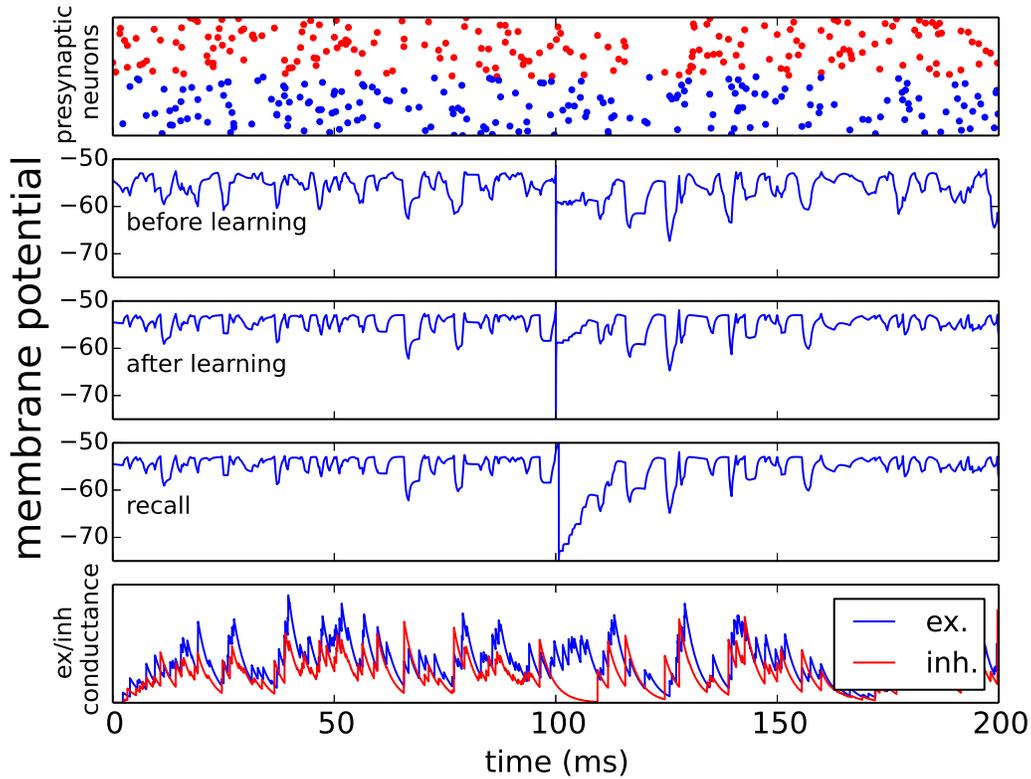}
 \end{center}
 \caption{{\bf Hebbian learning with homeostatic MPDP on inhibitory synapses.}
 A conductance based integrate-and-fire neuron is repeatedly presented with a
 fixed input pattern of activity in presynaptic inhibitory or excitatory neuron
 populations (top row - blue is excitatory, red inhibitory). Before learning,
 the neuron is allowed to adapt it's inhibitory weights according to
 homeostatic MPDP, such that the membrane potential mostly stays between the
 two learning thresholds. Then a strong excitatory input is given concurrently
 with the pattern to induce a spike at $t=100ms$ (second row). Learning is
 restricted to inhibitory weights. By release of inhibition, the net input
 after the teacher spike is increased (third row). After learning has
 converged, the neuron is presented the input pattern without the teacher input
 and reproduces the spike close to the target time (4th row) . At all other times,
 excitatory and inhibitory conductances are balanced (bottom row).}
 \label{fig2}
\end{figure}

To demonstrate the capability of MPDP for learning of exact spike times, we
constructed a simple yet plausible feed-forward network of $N_i$ inhibitory and
$N_e$ excitatory neurons. Synaptic weights were initialized randomly. Both
populations projected onto one conductance based LIF neuron. We presented this
network frozen poissonian noise as the sole presynaptic firing pattern (Fig.
~\ref{fig2}, top). Excitatory synapses were kept fixed and
inhibitory synapses changed according to MPDP. First we let the network learn to
balance all inputs from the excitatory population such that the membrane
potential mostly stays between the thresholds $\vartheta_P^I$ and
$\vartheta_D^I$. We then introduced the teacher input as a strong synaptic input
from a different source (e.g. a different neuron population, Fig.~\ref{fig2},
second to top). After repeated presentations of the input
pattern with the teacher input, inhibition around the teacher spike is released
such that after learning the output neuron will spike close to the desired
spike time even without the teacher input (Fig.~\ref{fig2}, third and fourth to top). At the same time,
due to the balance requirement of the learning rule, inhibitory and excitatory
conductances covary and thus their influence on the membrane potential mostly
cancels out (Fig.~\ref{fig2} bottom). Due to the
sterotypical shape of the membrane potential around the teacher spike, a
homeostatic learning rule is able to perform associative learning by release of
inhibition.

\begin{figure}[!ht]
 \begin{center}
  \includegraphics[width=0.95\textwidth]{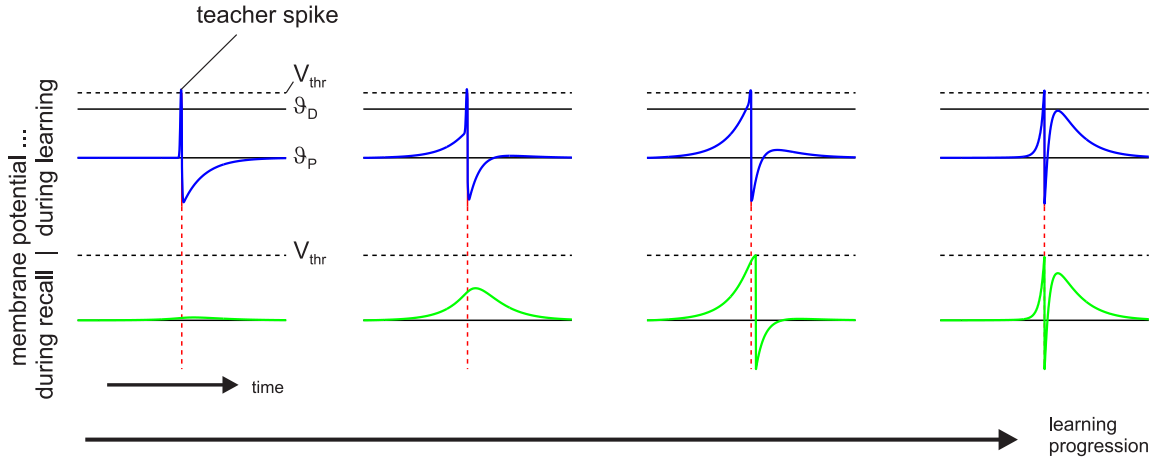}
 \end{center}
 \caption{{\bf Hebbian learning with homeostatic MPDP}. A postsynaptic neuron
 is presented the same input pattern multiple times, alternating between
 teaching trials with teacher spike (blue trace) and recall trials (green
 trace) to test the output. Initially, all weights are zero (left). Learning is
 Hebbian initially until strong depolarization occurs (second to left). When
 the spike first appears during recall, it is still not at the exact location
 of the teacher  spike (second to right). Continued learning moves it closer to
 the desired  location. Also, the time windows of the voltage being above
 $\vartheta_D$ and  below $\vartheta_P$ shrink and move closer in time (right).
 Synaptic plasticity almost stops. The number of learning trials before each
 state is 1, 16, 53, and 1600 from left to right.}
 \label{fig3}
\end{figure}

To further investigate the learning process, we simplified the setup. All
synapses were subject to MPDP and were allowed to change their sign.
A population of $N$ presynaptic
neurons fires one spike in each neuron at equidistant times. They project onto
a single postsynaptic LIF neuron and all weights are zero
initially. In each training trial an external delta-shaped suprathreshold
current is induced at the postsynaptic neuron at a fixed time relative to the
onset of the input pattern (teacher spike). The postsynaptic neuron reaches its
firing threshold instantaneously, spikes and undergoes reset into a
hyperpolarized state (blue trace on the left in Fig.~\ref{fig3}). This is mathematically equivalent to adding a
reset kernel at the time of the external current~\cite{Memmesheimer2014}.
Because we set $\vartheta_P = V_{eq} = 0$, potentiation is induced in all
synapses which have temporal overlap of their PSP-kernel with the hyperpolarization.
Probing the neuron a second time without the external spike shows a small bump
in the membrane potential around the time of the teacher spike. We continued to
present the same input pattern, alternating between teaching trials (with
teacher spike) and recall trials without teacher and with synaptic plasticity
switched off. Plasticity is Hebbian until the weights are strong enough such
that there is considerable depolarization before the teacher spike, inducing
synaptic depression. Also, spike
after-hyperpolarization is partially compensated by excitation, which reduces
the window for potentiation. Continuation of learning after the spike
association has been achieved (second to right plot) shrinks the windows for
depression and potentiation, until they are very narrow and very close to each
other in time. Because synaptic plasticity is determined by the integral over the
normalized PSP during periods of depolarization and hyperpolarization, depression
and potentiation become very similar in magnitude for each synapse and
synaptic plasticity slows down nearly to a stop. Furthermore, the output spike
has become stable. The time course of the membrane potential during teaching
and recall trials is almost the same (Fig.~\ref{fig3} right).

\subsection*{Quantitative evaluation of MPDP}

\subsubsection*{Memory capacity}

\begin{figure}
 \begin{center}
  \includegraphics[width=0.7\textwidth]{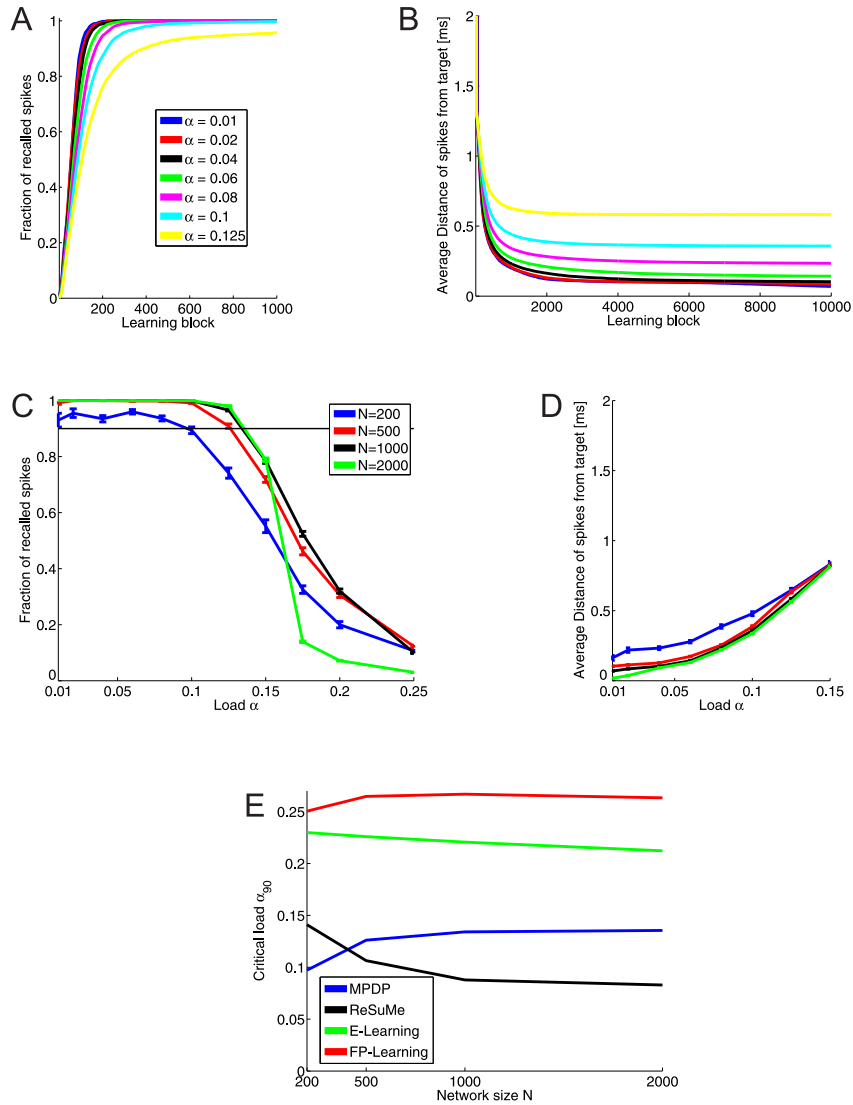}
 \end{center}
 \caption{{\bf Capacity of networks with MPDP.} {\bf A:} Fraction of pattern
 where the network generates an output spike within 2 ms distance of target time
 $t_d^{\mu}$, and no spurious spikes. Network size is $N = 1000$. The desired
 spikes are learned within $\approx$ 600 steps. {\bf B:} Average distance of
 output spikes to target for the same network size. Training continues even
 though the desired spikes are generated; however, they are pushed closer to
 the desired time. {\bf C:} Average fraction of recalled spikes after 10000
 learning blocks for all network sizes as a function of the load. Networks with
 $N = 200$ have a high probability to not be able to recall all spikes even for
 low loads. Otherwise, recall gets better with network size. The thin black
 line lies at fraction of recall equal to 90 \%. The critical load
 $\alpha_{90}$ is the point where the graph crosses this line. {\bf D:} Average
 distance of recalled spikes as a function of the load. The lower the loads,
 the closer the output spike are to their desired location. {\bf E:} Critical
 load as a function of network size for all four learning rules. MPDP reaches
 approximately half of the maximal capacity.}
 \label{fig4}
\end{figure}

We numerically evaluated the capacity of MPDP to train a network to produce precise
spike times using the simplified feed-forward network described above. We
constructed input patterns and desired output using the Chronotron framework
~\cite{Florian2012}. During training, we monitored the success of recall over
time. The network of size $N = 1000$ generates the desired output spikes within
the window of tolerance after 600 learning blocks (Fig.
~\ref{fig4} A). However, weights are still changed by
training, and continuation of it reduces the difference of actual and desired
output spike time (see Fig.~\ref{fig4} B). After around
2000 learning blocks the average temporal error of all recalled spikes stays constant for the
remainder of training. For
$\alpha \leq 0.1$ the self-generated output spike is on average less than 0.5 ms
away from the desired time. The final fraction of recalled spikes and average
distance are shown in Fig.~\ref{fig4} C and D.
The smallest network ($N = 200$) never reaches perfect recall, but has a
capacity of $\alpha_{90} = 0.095$ (for the definition of capacity, see Materials and Methods). All other networks
achieve perfect recall up to a load of $\alpha = 0.1$ and a capacity of 
$\alpha_{90} \approx 0.135$. The average distance of spikes from teacher grows
with the load, but stays below 0.5 ms.

To put these results into perspective, we trained Chronotrons again using three
other learning rules and computed the respective memory capacity. Fig.
~\ref{fig4} shows the capacity of all plasticity rules.
The upper bound established by FP-Learning is $\alpha_{90} \approx 0.26$.
MPDP is capable of storing half of the maximal possible number of associations
in the weights.

\subsubsection*{Training and recall with noise on the membrane potential}

\begin{figure}[ht]
 \begin{center}
  \includegraphics[width=0.7\textwidth]{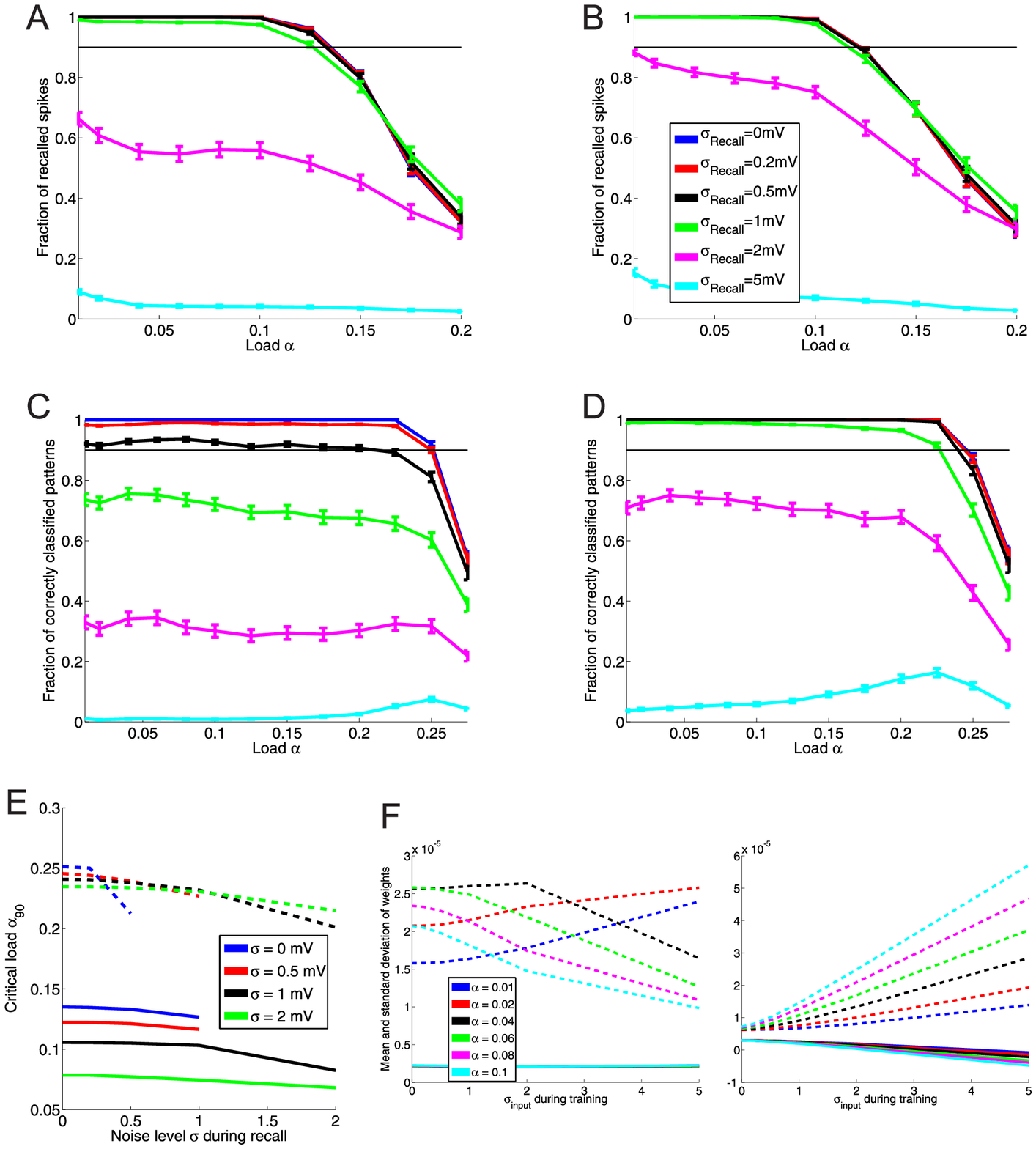}
 \end{center}
 \caption{{\bf Capacity of networks under input noise.} {\bf A:} Recall as a
 function of the load for different levels of noise during recall. Noise is
 imposed as an additional stochstastic external current. Networks were trained
 with MPDP. Up to a noise level $\sigma{input} = 1 mV$ during recall, there is
 almost no degradation of capacity. {\bf B:} Same as A, but with stochastic
 input noise of width $0.5 mV$ during network training. The capacity is
 slightly reduced, but resistance against noise is slightly better. {\bf C} and
 {\bf D:} Same as A and B, but the network was trained with FP-Learning. The
 capacity is doubled. However, the network trained without noise shows an
 immediate degradation of recall with noise. If the network is trained with
 noisy examples (D, $\sigma_{input} = 0.5 mV$), also recall with noise of the
 same magnitude is perfect. {\bf E:} Comparison of capacity of networks trained
 with MPDP and FP-Learning depending on input noise during training and recall.
 Solid lines: MPDP, dashed lines: FP-Learning. Lines that are cut off indicate
 that the network failed to reach 90 \% recall for higher noise. x-axis is
 noise level during recall. Different colors indicate noise level during
 training. Curiously, although FP-Learning suffers more from higher noise
 during recall than during training, the capacity drops less than with MPDP.
 {\bf F:} Comparison of weight statistics of MPDP (left) and FP-Learning
 (right) after learning. Solid lines are mean, dashed lines are standard
 deviation. With MPDP, the weigths stay within a bounded regime, the mean is
 independent of noise or load during training; the cyan line for $\alpha = 0.1$
 occludes the others. FP-Learning rescales the weights
 during training with noise: The mean becomes negative, and the standard
 deviation grows linearly with noise level. This effectively scales down the
 noise by stochastic input.}
 \label{fig5}
\end{figure}

Next, we turned to an evaluation of memory under the influence of noise. Having
a noise free network is a highly idealized situation and neurons in the brain
are more likely to be subject to noise, be it because of inherent stochasticity
of spike generation or the fact that sensory inputs are almost never ``pure'',
but likely to arrive with additional more or less random inputs. First, we tested
training and recall of spike times using an additional random input of a given
variance $\sigma_{input}$ on the postsynaptic neuron. The random input is a gaussian white noise
process with zero mean, and because inputs decay with the membrane time
constant, this results in a additional random walk with a restoring force. We trained the
Chronotron with additional noise of width $\sigma_{input} \in \{0, 0.2, 0.5, 1,
2, 5\} mV$. The width is the standard deviation of the random walk. Afterwards, we evaluated the critical load of networks of size $N =
200, 500, 1000$ depending on the noise level during training and during recall.
The results are shown in Fig.~\ref{fig5}.

With MPDP, the network trained without noise can perfectly recall patterns up
to a load of $\alpha = 0.1$ even with additional noise input of $\sigma_{input}
= 0.5 mV$. Adding noise during training decreases the capacity, but at the same
time recall robustness against noise is improved. This is contrasted by the
network trained with FP-Learning. Here, noise-free training results in a
network with imperfect recall under noise. However, noise during
training alleviates this problem. Training with a given noise width
$\sigma_{input}$ makes recall with the same and less noise width perfect. One
interesting observation is that unlike with MPDP, with FP-Learning the memory
capacity for noise-free recall stays constant regardless of noise during
training. This is explained by the variance of the weights after training. With
FP-Learning, the variance increases approximately linearly with noise width, while the mean of the weights 
decreases linearly into negative values. The resulting membrane
potential is strongly biased towards hyperpolarized states. What FP-Learning
effectively does during training is to scale down the noise relative to the
weights. This reduces the influence of noise, but also leads to a membrane
potential that stays below resting potential most of the time during input
activity. Because of the threshold for LTP, MPDP can not scale the weights
freely, therefore it suffers from a declining memory capacity.

\subsubsection*{Training and recall with input spike time jitter}

As a second noise condition we tested training and recall in the case that the
input spike times are not fixed. In each pattern presentation, we added to each
presynaptic spike time some random number drawn from a gaussian distribution
with mean zero and some given variance. The input is not frozen noise anymore,
but a jittered version of the underlying input pattern $\{ t_i^{\mu} \}$.
Similarly to the condition of a stochastic input current, we tested the capacity
of the network if during recall the input pattern are jittered or if during
training the input is jittered (but noise free during recall). 

\begin{figure}
 \begin{center}
  \includegraphics[width=0.8\textwidth]{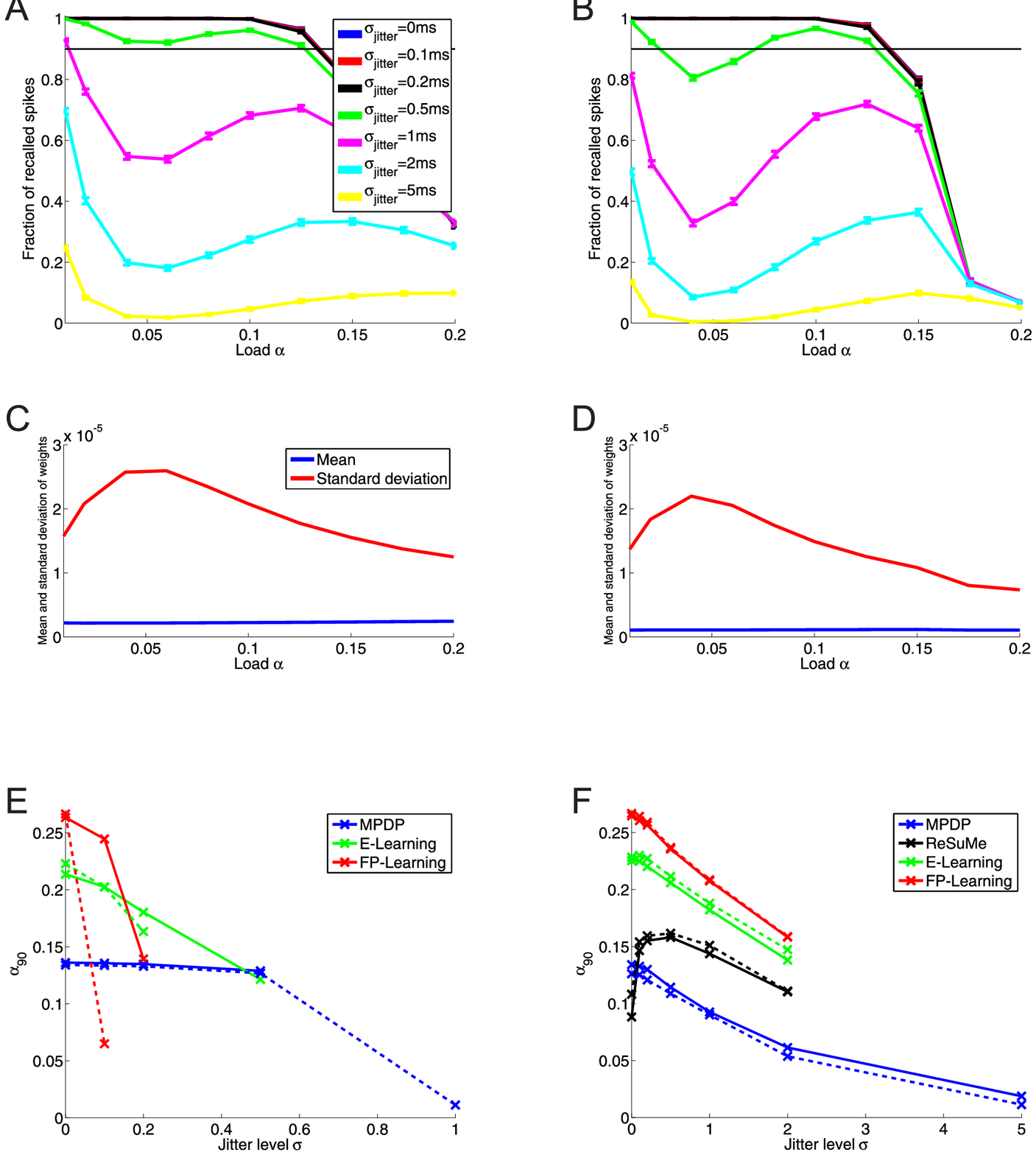}
 \end{center}
 \caption{{\bf Recall and capacity with input jitter.} {\bf A:} Recall of
 networks trained noise-free with MPDP if during recall the input patterns are
 jittered ($N = 1000$). The black line lies on top of the blue and red ones
 (same in B). Up to $\sigma_{jitter} = 0.5 ms$, the recall is
 unhindered. A curious feature is a ``slump'' in the recall for strong input
 jitter and intermediate loads. This slump is even more visible for the larger
 network with $N = 2000$ ({\bf B}). The slump strongly correlates with the
 variance of the weights as a function of network load ({\bf C} for $N = 1000$,
 {\bf D} for $N = 2000$). The mean of the weights stays almost constant. {\bf E:} Critical load as a function of input
 jitter during recall. The networks are trained noise free with different
 learning rules. Solid lines show $N = 2000$, dashed lines $N = 1000$. Crosses show sampling points. If a
 line is discontinued, this means that for this input jitter the networks do
 not reach 90 \% recall anymore. Recall for MPDP stays almost constant until
 $\sigma_{jitter} = 0.5$, while for the other learning rules a considerable 
 drop-off of recall is visible. {\bf F:} Noise free recall of networks trained
 with noisy input. For MPFP, E-Learning and FP-Learning alike the capacity
 drops with increasing training noise. The exception is ReSuMe. Here, the
 capacity strongly \emph{increases} if the noise is small.}
 \label{fig6}
\end{figure}

Fig.~\ref{fig6} A ($N = 1000$) and B ($N = 2000$) shows the
recall of networks trained noise free with MPDP if during recall
the spike times of the input patterns are jittered. For jitter with
a small variance ($\sigma_{jitter} < 0.5 ms$), the recall is almost
unaffected. For stronger jitter, recall deteriorates. A rather strange feature
of the recall is that for intermediate loads $\alpha \approx 0.05$ the recall
is worse than for loads close to the maximal capacity ($\alpha_{90}
\approx 0.125$). This observation is counter-intuitive and calls for explanation, because
recall usually becomes worse for memory systems if their load is close to the capacity.
However, fluctiations of the membrane potential due to jitter in the input
spike times are scaled by the weights. This seperates this noise condition from
the one with stochastic input current. A comparison of the weight statistics
of networks trained with MPDP after training shows that the slump in the recall
covaries with the weight variance (Fig.~\ref{fig6} C and
D). For $N = 1000$ the minimum of the slump lies at $\alpha = 0.06$, which coincides
with the maximum of the weight variance. For $N = 2000$, both lie at $\alpha
= 0.04$ instead. The mean of the weights does have little to no influence on
that; it stays almost constant as a function of load. E-Learning and
FP-Learning do not have the same characteristics (data not shown). For example, with
FP-Learning weight average and variance stay basically constant until a load
of $\alpha \approx 0.2$, rather close to the capacity. Only then the mean
decreases and variance increases (see for example Fig.~\ref{fig5} F, right plot for $\sigma_{input} = 0$ during
training).

Networks trained without noise and tested with jittered input show a similar
behavior to noise induced by an external stochastic current (Fig.~\ref{fig5} E,
blue lines, versus Fig.~\ref{fig6} E). Networks trained with MPDP tolerate noise up
to a certain degree without showing a deterioration of recall. With the other
learning rules, the recall gets worse with arbitrary small noise levels. On the other
hand, training a network with FP-Learning while injecting stochastic currents
(the previous noise condition) led to almost unharmed capacity. The reason is
that FP-Learning ``downscales'' the noise by scaling up the weight variance.
This is not a viable path for jitter of input spike times. Therefore,
E-Learning and FP-Learning as well as MPDP show a decrease of capacity if during
training the input spike times are jittered. An interesting outlier is ReSuMe.
The networks trained noise free with ReSuMe have low capacity and unstable recall. Even
with slight jitter the recall does not reach 90 \% anymore. Therefore, we do not
include ReSuMe in Fig.~\ref{fig6} E. However, training
the network with jitter leads to an increase of capacity (Fig.~\ref{fig6} F).

\section*{Discussion}

We introduced a synaptic plasticity mechanism that is based on the requirement
to balance the membrane potential and therefore uses the postsynaptic
membrane potential rather than postsynaptic spike times as the relevant signal
for synaptic changes (Membrane Potential Dependent Plasticity, MPDP). We have
shown that this simple rule allows the somewhat paradoxical temporal association
of enforced output spikes with arbitrary frozen noise input spike patterns (Chronotron).
Before, this task could only be achieved with supervised learning rules that
provided knowledge not only about the desired spike times, but also about the type of
each postsynaptic spike (desired or spurious). With MPDP, the supervisor only has to provide
the desired spike, while the synapse endowed with MPDP distinguishes between
desired and spurios spikes exploiting the time course of the voltage around the
spike. Additionally, the sensitivity of MPDP to subthreshold membrane potential
allows for robustness against noise.

\subsection*{Biological plausibility of MPDP}

Spike-Timing-Dependent Plasticity (STDP) is experimentally well established and
simple to formalize, which made it a widely used plasticity mechanism in
modelling. It is therefore important to note that MPDP is compatible with
experimental results on STDP, in particular with those of Hebbian STDP on inhibitory
synapses. The reason is that spikes come with a stereotypic trace in the membrane potential. The
voltage rises to the threshold, the spike itself is a short and strong
depolarization, and afterwards the neuron undergoes reset, all of which are
signals for MPDP. Pairing a postsynaptic spike with presynaptic spikes at
different timings gives rise to a plasticity window which shares its main
features with the STDP window: The magnitude of weight change drops with the
temporal distance between both spikes and the sign switches close to
concurrent spiking.

It is known that the somatic membrane potential plays a role in synaptic
plasticity. Many studies investigated the effect of prolonged voltage
deflections by clamping the voltage for an extended time while repeatedly
exciting presynaptic neurons (e.g. see~\cite{Artola1990}). However, MPDP predicts that synaptic plasticity is
sensitive to the exact time course of the membrane potential, as well as the
timing of presynaptic spikes. This necessitates that dendrites and spines
reproduce the time course of somatic voltage without substantial attenuation.
Morphologically the dendritic spines form a compartement separated from the
dendrite, which, for example, keeps calcium localized in the spine. It has been a
topic under investigation whether the spine neck dampens invading currents.
Despite experimental difficulties in measuring spine voltage, recent studies
found that backpropagating action potentials indeed invade spines almost
unhindered~\cite{Holthoff2010}. Furthermore, independently of spine morphology
and proximity to soma, the time course of a somatic hyperpolarizing current
step is well reproduced in dendrites~\cite{Palmer2009} and spines~\cite{Popovic2014}. This shows that at least
in principle the somatic voltage trace can be available at the synapse. In turn,
voltage-dependent calcium channels can transform subthreshold voltage
deflections into an influx of calcium, the major messenger for synaptic
plasticity. A few studies found that short depolarization events act as signals
for synaptic plasticity~\cite{Sjostrom2004, Fino2009}, with a dependence of
sign and magnitude of weight change on the timing of presynaptic spikes.

Another important point is the sign of synaptic change. ``Membrane Potential
Dependent Plasticity'' per se is a very general term which potentially could
include many different rules~\cite{Clopath2010, Shouval2002}. In this study, MPDP serves
as a mechanism that keeps the membrane potential bounded. For
inhibitory synapses this requirement results in a Hebbian plasticity rule,
which has been reported previously~\cite{Haas2006}.
Inhibitory neurons in cortex have been implied to precisely balance excitatory inputs
~\cite{Xue2014}. MPDP on excitatory synapses is necessarily
``Anti-Hebbian''. Lamsa et al~ \cite{Lamsa2007} found that pairing presynaptic
spikes with postsynaptic hyperpolarization can lead to synaptic potentiation.
This was caused by calcium permeable AMPA receptors (CP-AMPARs) present in
these synapses. However, Anti-Hebbian plasticity does not rely on
CP-AMPARs alone. Verhoog et al.~\cite{Verhoog2013}, for example, found
Anti-Hebbian STDP in human cortex, which depends on dendritic voltage-dependent
calcium channels. Taken together, these findings demonstrate the existence of
cellular machinery which could implement
homeostatic MPDP, either on excitatory or inhibitory synapses.

\subsection*{Properties and capabilities of Homeostatic MPDP}

We derived homeostatic MPDP from a balance requirement: Synapses change in
order to prevent hyperpolarization and strong depolarization for recurring
input activity. This kind of balance reduces metabolic costs of a neuron and 
keeps it at a sensible and sensitive point of operation~\cite{Attwell2001}. The
resulting plasticity rule is Anti-Hebbian in nature because synapses change to
decrease net input when the postsynaptic neuron is excited and to increase net
input when it is inhibited. However, spike after-hyperpolarization turns
homeostatic MPDP effectively into Hebbian plasticity. Every postsynaptic spike
causes a voltage reset into a hyperpolarized state. Therefore synapses of
presynaptic neurons which fired close in time to the postsynaptic spike will
change to increase net input if the same spatio-temporal input pattern
re-occurs. The total change summed over all synapses depends on the duration
and magnitude of hyperpolarization. Because the induced synaptic change reduces
this duration, total synaptic change is also reduced. The same is true for
total synaptic change to decrease net input, which depends on the duration
where the membrane potential stays above $\vartheta_D$ (resp. $\vartheta_P^I$
for inhibitory synapses) and which reduces this duration in future occurances.
If the rise time of the voltage before the spike and residual spike
after-hyperpolarization are both short and close in time, potentiation and
depression will become approximately cancelled around a spike.

In this view, associative synaptic plasticity or ``learning'' is the
consequence of imbalance. A spike is stable if the time course of the voltage
in its proximity leads to balanced weight changes. For example, if input is
just sufficient to cause a spike, the voltage slope just before the spike is
shallow and synaptic depression outweighs potentiation. On the other hand, the
delta-pulse shaped currents used to excite the postsynaptic neuron during
Chronotron training are very strong inputs. They are not unlearned. Instead,
the weights potentiate until the membrane potential is in a balanced state, and
the neuron fires the teacher spike on its own when left alone.

Another interesting aspect of MPDP is the emergence of robustness against noise.
Most obviously, with the choice of the threshold for depression the neuron sets
a minimal distance of the voltage to the firing threshold for known input
patterns. This allows to have perfect recall in the case of noisy input in the
Chronotron. The second effect of the depression threshold is more subtle. Not
only does it prevent spurious spikes, but through learning the slope of the
membrane potential just before the desired spike tends to become steep. This is
necessary to prevent spike extinction by noise. To see how this influences noise robustness, consider an output spike with a
flat slope of the voltage. Increasing the voltage slightly around the spike
time moves the intersection of the voltage with the firing threshold forward in
time by a proportionally large margin. Decreasing voltage moves it backwards in
time or could even extinguish the spike; a flat slope implies a low peak of the
``virtual'' membrane potential. MPDP in contrast achieves a state which is
robust against spike extinction as well as the generation of spurious spikes.
On the downside, keeping the voltage away from the firing threshold as well as
imposing steepness on the slope just before spikes puts additional constraints
on the weights. Robustness comes at the cost of capacity.

\subsection*{Relation of MPDP to other learning rules}

There are many supervised learning algorithms that are used to train neuronal
networks to generate desired spatio-temporal activity patterns. All of them
involve a comparison of the self-generated output to the desired target
activity. They can be broadly put into three different classes. E-Learning and
FP-Learning~\cite{Florian2012, Memmesheimer2014} are examples of algorithms of
the first class which are used to train a neuron to generate spikes at exactly
defined times. They first observe the complete output and then evaluate it
against the target. E-Learning performs a gradient descent on the
Victor-Purpura distance~\cite{Victor1996} between both spike trains. This means
that the weight changes associated to one particular spike (actual or desired)
can depend on distant output spikes. In FP-Learning, the training trial is
interrupted if the algorithm encounters an output error. Subsequent spikes are
not evaluated anymore. Thereby these algorithms are non-local in time and very artificial.

Another class of learning algorithms emerged recently with the examples PBSNLR~\cite{Xu2013} and HTP
~\cite{Memmesheimer2014}. They take an entirely different route. The
postsynaptic membrane potential is treated as a static sum of PSP kernels
weighted by the respective synaptic weight, similar to the SRM${}_0$ model of
the LIF neuron. The firing threshold is moved towards infinity to prevent
output spikes and voltage resets are added at the target spike times. Then the
algorithms perform a perceptron classification on discretely sampled time
points of the voltage, with the aim to keep it below the actual firing
threshold for all non-spike times and to make sure a threshold crossing at the
desired spike times. These algorithms were devised as purely
technical solutions and are highly artificial. However,
MPDP bears some similarity to the described procedure: Except close to teacher
inputs, at every point in time recently active synapses get depressed if the
voltage is above the threshold for depression. This is comparable to a
perceptron classification on a continuous set of points.

A third class of algorithms compares actual and target activity locally in
time. In contrast to the algorithms mentioned above, they are usually not used
to learn exact spike times, but rather continuous time dependent firing rates.
The ur-example is the Widrow-Hoff rule~\cite{HKP, Ponulak2010}. More recently,
similar rules were developed by Xie and Seung \cite{Xie2004}, Brea et al.
~\cite{Brea2013} and Urbanzcik and Senn~\cite{Urbanczik2014}. In contrast to the
Widrow-Hoff rule, the more recent rules are defined for spiking LIF neurons with a
``soft'' firing threshold, i.e. spike generation is stochastic and the
probability of firing a spike is a monotonous function of the current voltage.
In these rules, at every point in time the synaptic change is proportional to
the difference of the current firing rate and a target firing rate specified by
an external supervisor. When it comes to biological implementation, the central
problem of Widrow-Hoff type rules is the comparison of self-generated
and target activity. It is derived from the abstract goal to imprint the target
activity into the network. This target needs to be communicated to the neuron
and synaptic plasticity has to be sensitive to the difference of the neurons' own
current acticity state (implicitely represented by its membrane potential) and
the desired target activity. Usually, no plausible biological implementation
for this comparison is given. The combination of homeostatic MPDP,
hyperpolarization and a teacher now offers a solution to both problems. The
teacher provides information about the target activity through temporally
confined, strong input currents which cause a spike. Spike
after-hyperpolarization (SAHP) allows to compare the actual input to the target
without inducing spurious spikes detrimental to learning. The more SAHP is
compensated by synaptic inputs, the closer the self-generated activity is to
the target and the less synapses need to be potentiated. This is implemented
naturally in MPDP, where potentiation is proportional to the magnitude and
duration of hyperpolarization. On the other hand, strong subthreshold
depolarization implies that self-generated spurious spikes are highly probable,
and weights need to be depressed to prevent spurious spikes in future
presentations.

A further solution for the problem of how information about the target is
provided was given by Urbanczik and Senn~\cite{Urbanczik2014}. Here, the neuron
is modelled with soma and dendrite as seperate compartements instead of point neurons as used in this study. The
teacher is emulated by synaptic input projecting directly onto the soma, which causes a specfic time
course of the somatic membrane potential. The voltage in the dendrite is
determined by a different set of synaptic inputs, but not influenced by the
somatic voltage; however, the soma gets input from the dendrites. The
weight change rule then acts to minimize the difference of somatic (teacher)
spiking and the activity as it would be caused by the current dendritic voltage.
This model represents a natural way to introduce an otherwise abstract teacher
into the neuron. Nonetheless, the neuron still has to estimate a firing rate from
its current dendritic voltage, for which no explicit synaptic mechanism is
provided. Also, it is worth noting that the model of Urbanczik and Senn
requires a one-way barrier to prevent somatic voltage invading the dendrites;
in contrast, MPDP requires a strong two-way coupling between somatic and dendritic/synaptic
voltage.

Another putative mechanism for a biolgical implementation of the
$\delta$-rule was provided by D'Souza et al.~\cite{DSouza2010}. In this model, a neuron
recieves early auditory and late visual input. By the combination of spike
frequency adaptation (SFA) and STDP, the visual input acts as the teacher that
imprints the desired response to a given auditory input in an associative
manner. However, the model is quite specific to the barn owl setting; for
example, parameters have to be tuned to the delay between auditory and visual
input.

Applying rules of the Widrow-Hoff type to fully deterministic neurons can lead
to unsatisfactory results. ReSuMe is an example of such a rule~\cite{Ponulak2010}. Its memory capacity is low,
but it increases sharply if the input is noisy during training (see Fig.
\ref{fig6}). A propable reason is that in a fully
deterministic setting, the actual spike times do not allow a good estimation of
the expected activity. This sounds paradoxial. But if we consider a deterministic
neuron with noise-free inputs the membrane potential can be
arbitrarily close to the firing threshold without crossing it. But even
the slightest perturbation can cause spurious spikes at those times.
This leads to bad convergence in Chronotron training, since the perturbations
caused by weight changes for one pattern can easily destroy previously learned correct
output for another pattern~\cite{Memmesheimer2014}.
The problem of these rules is the sensing of the activity via the instantaneous firing rate.
Therefore, the explicit sensitivity to subthreshold voltages of MPDP is advantageous if
training examples are noise free.

We conclude that our MPDP rule with hyperpolarization and teacher input
represents a biologically plausible implementation of the comparison of actual
and target activity that is key to all supervised learning algorithms. Also,
because MPDP is explicitely sensitive to the membrane potential and not the
firing rate, it is fully applicable to deterministic neurons. Additionally, the
training procedure leads to networks whose output is robust against input
noise, similar to what learning algorithms of the Widrow-Hoff type achieve.

\subsection*{Outlook}

We derived the synaptic plasticity rule from the objective to keep the 
membrane potential within bounds, which is a homeostatic principle that 
at first sight would primarily serve the stability of network dynamics. 
In particular, this principle might explain the strikingly detailed 
balance of excitation and inhibition as observed in cortex~\cite{Wehr2003, Haider2006, Okun2008}
(compare also Fig.~\ref{fig2}, bottom row). In fact, such homeostatic plasticity 
has been found e.g. for parvalbumin expressing interneurons which 
selectively adapt their synaptic strength in an activity dependent 
manner to match the excitatory inputs to target cells~\cite{Xue2014}. Being an 
anti-hebbian mechanism homeostatic plasticity might even appear to contradict associative 
learning. Therefore we find it particularly intriguing that -when 
combined with the ubiquitous spike after-hyperpolarizarion- it can 
paradoxically entail robust spike-based associative learning. We think 
this fact suggests that the balance in cortex could rather reflect a
powerful learning principle at work.

\bibliography{Chronotron_paper_v5}

\end{document}